\documentclass{emulateapj}
\usepackage{epsfig}

\newcommand{\Lx}{\mbox{$L_{\rm X}$}}
\newcommand{\Tx}{\mbox{$T_{\rm X}$}}
\newcommand{\etal}{et al.{}}
\newcommand{\LCDM}{{\mbox{$\Lambda$CDM}}}
\newcommand{\Msun}{{\, \mbox{M$_\odot$}}}
\newcommand{\hMsun}{\, h^{-1} \Msun}

\newcommand{\Mpc} {{\rm {Mpc}}}

\newcommand{\hMpc}{\mbox{$\, h^{-1}\Mpc$}}

\newcommand{\spose}[1]{{\hbox to 0pt{#1\hss}}}
\newcommand{\lta}{\mathrel{\spose{\lower 3pt\hbox{$\mathchar"218$}}
     \raise 2.0pt\hbox{$\mathchar"13C$}}}
\newcommand{\gta}{\mathrel{\spose{\lower 3pt\hbox{$\mathchar"218$}}
     \raise 2.0pt\hbox{$\mathchar"13E$}}}
\newcommand{\approxlt}{\mathrel{\spose{\lower 3pt\hbox{$\sim$}}
        \raise 2.0pt\hbox{$<$}}}
\newcommand{\approxgt}{\mathrel{\spose{\lower 3pt\hbox{$\sim$}}
        \raise 2.0pt\hbox{$>$}}}
\newcommand{\approxpropto}{\mathrel{\spose{\lower 3pt\hbox{$\sim$}}
        \raise 2.0pt\hbox{$\propto$}}}
\newcommand{\twiddles}[1]{\spose{\raise 5pt\hbox{$\sim$}}\hbox{$#1$}}

\shorttitle{Evolution of X-ray clusters} 
\shortauthors{Muanwong \etal}

\begin{document}

\title{Evolution of X-ray cluster scaling relations in
simulations with radiative cooling and non-gravitational heating}

\author{Orrarujee Muanwong\altaffilmark{1},
        Scott T. Kay\altaffilmark{2,3} and
        Peter A. Thomas\altaffilmark{3}}

\altaffiltext{1}{Department of Physics, Faculty of Science, 
Khon Kaen University, Khon Kaen, 40002, Thailand; orrmua@kku.ac.th.}
\altaffiltext{2}{Astrophysics, Denys Wilkinson Building, Keble Road,
  University of Oxford, Oxford, OX1\,3RH,
  UK.}
\altaffiltext{3}{Astronomy Centre, Department of Physics and Astronomy, 
School of Science and Technology, University of Sussex, Falmer, Brighton,
  BN1\,9QH, UK.}

\begin{abstract}
We investigate the redshift dependence of X-ray cluster scaling
relations drawn from three hydrodynamic simulations of the
$\Lambda$CDM cosmology: a {\it Radiative} model that incorporates
radiative cooling of the gas, a {\it Preheating} model that
additionally heats the gas uniformly at high redshift, and a {\it
Feedback} model that self-consistently heats cold gas in proportion to
its local star-formation rate.  While all three models are
capable of reproducing the observed local \Lx-\Tx\ relation, they
predict substantially different results at high redshift (to $z=1.5$),
with the {\it Radiative}, {\it Preheating} and {\it Feedback} models
predicting strongly positive, mildly positive and mildly negative
evolution, respectively.

The physical explanation for these differences lies in the structure
of the intracluster medium.  All three models predict significant
temperature fluctuations at any given radius due to the presence of
cool subclumps and, in the case of the {\it Feedback} simulation,
reheated gas.  The mean gas temperature lies above th e dynamical
temperature of the halo for all models at $z=0$, but differs
between models at higher redshift with the {\it Radiative} model
having the lowest mean gaswos temperature at $z=1.5$.

We have not attempted to model the scaling relations in a manner that
mimics the observational selection effects, nor has a consistent
observational picture yet emerged.  Nevertheless, evolution of the
scaling relations promises to be a powerful probe of the physics of
entropy generation in clusters.  First indications are that early,
widespread heating is favored over an extended period of heating that
is associated with galaxy formation.
\end{abstract}


\section{Introduction}
\label{sec:introduction}

X-ray scaling relations of galaxy clusters, namely the temperature--mass,
\Tx-$M$, relation and the luminosity--temperature, \Lx-\Tx, relation, 
play a pivotal role when using the abundance of clusters to constrain 
cosmological parameters
\citep{HA91,WEF93,Eke96,VL96,VL99,Henry97,Henry00,Borgani01,Pierpaoli01,
Seljak02,Pierpaoli03,Viana03,Allen03,Henry04}. It is well known, however,
that accurate calibration of scaling relations is crucial to avoid a major
source of systematic error. 
For example, the \Tx$-M$ relation is widely used by many of these authors to 
constrain the amplitude of mass fluctuations, conventionally defined using
the parameter, $\sigma_8$. Systematic deviations in the normalization of the 
\Tx$-M$ relation, particularly due to how cluster mass is estimated 
(e.g. see \citealt{HMS99}) is amplified by the steep slope of the 
temperature function, leading to large variations
in $\sigma_8$ (see \citealt{Henry04} for a discussion of recent
results). 

As far as the \Lx-\Tx\ relation is concerned, the discrepancies are
more prominent as \Lx\ is highly sensitive to the thermodynamics of
the of the inner intracluster medium (ICM), and can yield different
values for both normalizations and slopes
\citep{EdS91,WJF97,AlF98,Mar98,XuW00}.  The situation is further
complicated by the fact that clusters do not scale self-similarly, as
would be the case (approximately) if the only source of heating was via
gravitational infall \citep{Kaiser86}. This makes the problem more
difficult to investigate theoretically, although it allows studies of
cluster scaling relations to reveal more information on the physics
governing the structure of the intracluster medium.

The departure from self-similarity can be attributed to an increase in
the {\it entropy} of the gas that particularly affects low-mass
systems \citep{EH91,Kaiser91,Bower97,TN01,PCN99,VB01,Voit02,Voit03}.
Many theoretical studies have been performed to investigate the
effects of various physical processes that can raise the entropy of
the gas, based on models involving heating
\citep{ME94,Balogh99,KY00,Low00,WFN00,Bower01,Borgani02} , radiative
cooling
\citep{KP97,Pearce00,Bryan00,Muanwong01,Muanwong02,DKW02,WX02}, and a
combination of the two
\citep{Muanwong02,KTT03,Tornatore03,Valdarnini03,Borgani04,Kay04,McCarthy04}.

Measurements of how cluster scaling relations evolve with redshift
allow even tighter constraints to be placed on cosmological parameters
(and entropy generation models), and observations of cluster
properties at high redshift are now starting to become available,
owing primarily to the high sensitivity of {\it Chandra} and {\it
XMM--Newton}. From a theoretical point of view, this is an exciting
phase as we can now fully exploit the availability of our simulated
distant clusters and compare their X-ray properties with real
observations. It is therefore timely to investigate further the
effects of entropy generation on the evolution of clu ster scaling
relations as the available data for high-redshift systems accumulates.

In this paper, we will use cosmological hydrodynamical 
simulations described in \citet{Muanwong02}, hereafter MTKP02, and
in \citet{Kay04}, hereafter KTJP04, to trace the
evolution of the cluster population to high redshift ($z=1.5$). Our
results will primarily focus on three ({\it Radiative}, {\it Preheating} 
and {\it Feedback}) models, all able to reproduce the local 
\Lx-\Tx\ relation. The aims of this paper are to determine how the scaling
relations evolve with redshift in the three models
and to discover what the evolution of scaling relations can 
teach us about non-gravitational processes occurring in clusters.

The rest of this paper is outlined as follows.  In
Section~\ref{sec:srel} we introduce the X-ray scaling relations and
summarize our present observational knowledge of these quantities.
Details of our simulated cluster populations are presented in
Section~\ref{sec:sims}.  In Section~\ref{sec:results} we present our
main results, first at $z=0$, where the models are in good agreement
with each other and the observations, then as a function of redshift,
where the models predict widely different results.  We discuss the
implications of these differences in Section~\ref{sec:discuss} and
demonstrate that the degree of X-ray evolution is driven by the supply
of cold, low entropy gas.  Finally, we summarize our conclusions in
Section~\ref{sec:conclude}.

\section{X-ray cluster scaling relations}
\label{sec:srel}

\citet{Kaiser86} derived the following relations for temperature
\begin{equation}
\Tx \propto M^{2 \over 3} \, (1+z),
\label{eqn:tmrel}
\end{equation}
and luminosity
\begin{eqnarray}
\Lx &\propto& M^{4 \over 3} \, (1+z)^{7 \over 2} \label{eqn:lmrel} \\
    &\propto& \Tx^{2} \, (1+z)^{3 \over 2} \label{eqn:ltrel},
\end{eqnarray}
assuming the distribution of gas and dark matter in clusters is
perfectly self-similar and the X-ray emission is primarily thermal
bremsstrahlung radiation. Observed clusters do not form a self-similar
population but it is nevertheless convenient to describe their
behavior using a generalized power-law form
\begin{equation}
Y = C_0(z) \, X^\alpha = Y_0 \,  X^\alpha \, (1+z)^{A},
\label{eqn:powerlaw}
\end{equation}
where $C_0(z)$ and $Y_0$ determine the normalization, $\alpha$ is the slope
of the relation (in log-space) and $A$ determines how  the relation evolves 
with redshift.  Our main results will focus on the determination of $A$.

\begin{figure}
\centerline{\epsfig{file=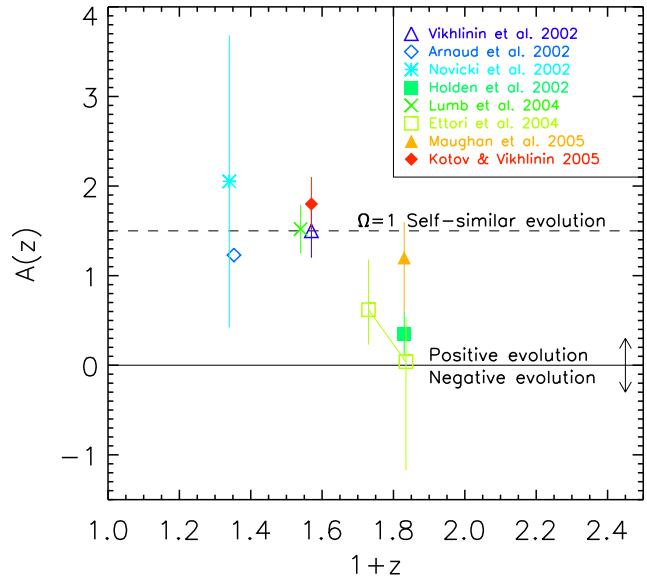,width=10cm,angle=0}}
\caption{Evolution in the \Lx-\Tx \ relation as measured from various
high redshift cluster samples.}
\label{fig:ltevolobs}
\end{figure}

Observationally, attempts to measure the \Tx$-M$ relation at high
redshift are currently in their infancy, as they require temperature
profiles to be measured so that their mass can be estimated, but
initial results are consistent with self-similar evolution ($A \sim
1$, \citealt{Maughan05,Kotov05}).

Measuring the \Lx-\Tx\ relation at higher redshift is a somewhat
simpler prospect, and has been attempted by many authors
\citep{MS97,Fairley00,Holden02,Novicki02,Arnaud02,Vikhlinin02,Lumb04,
Ettori04,Maughan05,Kotov05}. We summarize recent results that adopt a
low-density flat cosmology in Figure~\ref{fig:ltevolobs}, attempting
to include in the size of the error bars the uncertainty in $A$ due
to the choice of local relation (when quoted by the authors). Although
the present situation is by no means clear, taking all results at face
value generally favors positive evolution ($0 \approxlt A \approxlt
2$) with the latest results being consistent with sel f-similar
evolution ($A=3/2$). Larger samples of high
redshift clusters (such as that expected from the {\it XMM-Newton}
Cluster Survey, \citealt{Romer01}) will be crucial to accurately
constrain the degree of evolution in the \Lx-\Tx\ relation.

\section{Simulated cluster populations}
\label{sec:sims}

Our results are drawn from three similarly-sized $N$-body/SPH
simulations of the $\Lambda$CDM cosmology, which have already
been published in MTKP02 and KTJP04. The simulation box in MTKP02 has
a comoving side of $100\hMpc$ with $160^3$ particles each of gas and 
dark matter, whose particle masses are set to $2.6\times 10^9$ and
$2.1\times 10^{10}\hMsun$, respectively. The box used in KTJP04 is bigger 
with a side of $120\hMpc$ using $256^3$ particles each of gas and dark 
matter, whose particle masses are $1.3\times 10^9$ and $7.3\times 10^9\hMsun$,
respectively. Full details can be found in the articles. The key difference 
between the simulations is the model used to raise the entropy of 
the intracluster gas, summarized as follows:

\begin{enumerate}
\item A {\it Radiative} model where the excess entropy originated from
the removal of low entropy gas to form stars, causing higher entropy
gas to flow adiabatically into the core from larger radii (MTKP02).
\item  A {\it Preheating} model where entropy was generated impulsively
by uniformly heating the gas by 1.5 keV per particle at $z=4$ (MTKP02).
\item A {\it Feedback} model where the entropy of (on average) 
10 per cent of cooled gas in high density regions was raised by
1000 keV cm$^2$, mimicking the effects of heating due to stars and
active galactic nuclei (KTJP04).
\end{enumerate}

These three models differ in the timing and distribution of entropy
generation in the intracluster medium.  The {\it Radiative} model has no
explicit feedback of energy but relies on the removal of low-entropy
gas via cooling; as such it represents a minimal heating model.  The
{\it Preheating} model contains distributed heating at high redshift
such as
might occur if entropy generation occurs mainly in low-mass galaxies.
By contrast heating in the {\it Feedback} model occurs solely in
high-density regions. In all our models, there is
very little star formation before a
redshift of z=4 after which the star-formation rate (sfr) begins to
rise rapidly.  In the {\it Preheating} simulation the sfr is then
strongly
suppressed, whereas in the other two simulations it peaks at a redshift
of z=2 and then declines back down to low values by the present day
with a time-variation that matches that of the
star-formation history of the Universe.
The global baryon fraction in stars (and cold gas) at z=0 is 0.002,
0.076 and 0.127 in the {\it Preheating}, {\it Radiative} and
{\it Feedback} simulations,
respectively.  The largest of these corresponds to a stellar mass
density of $\Omega_*$=0.006; thus none of the models has excessive
star-formation.  These models are far from
exhaustive and their precise details should not be taken too
seriously.  The purpose of this paper is not to examine particular
models but to illustrate that the evolution of the X-ray scaling
relations can provide a powerful discriminant between different
classes of model.

\subsection{Cluster identification and properties}

Clusters were selected at four redshifts ($z=$0, 0.5, 1 \& 1.5) using the
procedure outlined in MTKP02.  They are defined to be spheres of
matter, centered on the dark matter density maximum, with total mass
\begin{equation}
M_{\Delta} = {4 \over 3} \pi R_{\Delta}^{3} \, \Delta \, \rho_{\rm c0}
\, (1+z)^{3},
\label{eqn:mass}
\end{equation}
where $\rho_{\rm c0}=3H_0^2/8\pi G$ is the critical density at $z=0$.
We set $\Delta=500$ as it corresponds to a sufficiently large radius
such that the results are not dominated by the core, as well as
corresponding approximately to the extent of current X-ray
observations. Furthermore, as was shown by \citet{Rowley04}, the X-ray
properties of simulated clusters within an overdensity of 500 exhibit
less scatter than within the virial radius.  Our choice of scaling
with redshift\footnote{Many authors prefer to adopt the redshift
scaling of the critical density, $E(z)^2=\Omega_{\rm m}(1+z)^{3} +
\Omega_{\Lambda}$ (for a flat universe), rather than the background
density, $(1+z)^3$.} is independent of cosmology and would allow the
simple power-law scalings to be recovered
(equations~\ref{eqn:tmrel},\ref{eqn:lmrel} \& \ref{eqn:ltrel}) if the
clusters were structurally self-similar.

We consider scaling relations involving mass, three measures of
temperature, and luminosity, for particle properties averaged within
$R_{500}$.  The mass,  
\begin{equation}
M_{500}=\sum_i m_i,
\end{equation}
where the sum runs over all particles, of mass $m_i$.
The dynamical temperature,
\begin{equation}
kT_{\rm dyn} = \, { \sum_{i,{\rm gas}} m_i k T_i \, + \, \alpha
\sum_i {1 \over 2} m_i v_i^2 \over \sum_i m_i},
\label{eqn:tdyn}
\end{equation}
where $\alpha=(2/3)\mu m_{\rm H} \sim 4.2\times 10^{-16}$ keV for a
fully ionized primordial plasma, assuming the ratio of specific heats
for a monatomic ideal gas, $\gamma=5/3$, and the mean atomic weight of
a zero metalicity gas, $\mu m_{\rm H}=10^{-24} {\rm g}$.  The first
sum in the numerator runs over all gas particles, of temperature,
$T_i$, whereas the second sum runs over particles of all types, of
speed $v_i$ as measured in the center of momentum frame of the
cluster.

We also consider the mass-weighted temperature of hot ($T>10^{5}$K) gas,
\begin{equation}
kT_{\rm gas} = { \sum_{i,{\rm hot}} m_i k T_i \over
\sum_{i,{\rm hot}} m_i},
\label{eqn:tgas}
\end{equation}
and we approximate the X-ray temperature of a cluster using the
bolometric emission-weighted temperature,
\begin{equation}
kT_{\rm bol} = { \sum_{i,{\rm hot}} m_i \rho_i \Lambda_{\rm
        bol}(T_i,Z) T_i \over \sum_{i,{\rm hot}} m_i \rho_i
        \Lambda_{\rm bol}(T_i,Z) },
\label{eqn:tbol}        
\end{equation}
where $\rho_i$ is the density and $\Lambda_{\rm bol}$ is the
bolometric cooling function used in our simulations \citep{SD93}; for
the {\it Radiative} and {\it Preheating} runs, $Z=0.3(t/t_0)Z_{\odot}$
(MTKP02), and for the {\it Feedback} run, $Z=0.3Z_{\odot}$
(KTJP04). 
Finally, the X-ray luminosity is approximated by the
bolometric emission-weighted luminosity
\begin{equation}
L_{\rm bol} =  \sum_{i,{\rm hot}} 
              {m_i \rho_i \Lambda_{\rm bol}(T_i,Z)
              \over (\mu m_{\rm H})^2 }.
\end{equation}

It has been shown recently that the emission-weighted temperature is
not an accurate diagnostic of cluster temperature, overpredicting the
{\it spectroscopic} temperature by $\sim 20-30$ per cent when the
emission is predominantly thermal bremsstrahlung
\citep{Mazzotta04,Rasia05}.  At lower temperatures ($kT<3$keV), line
emission from heavy elements makes the problem significantly more
complicated \citep{Vikhlinin05}.  The volume sampled by our
simulations ($\sim 100\hMpc$) means that we have very few clusters
with $T>3$keV, and so a more accurate measure of the cluster
temperature would require significantly more effort than applying a
simple formula to our data. We therefore leave such improvements to
future work, when larger samples of simulated clusters are available.
It would not affect the conclusions of this paper.

\subsection{Cluster catalogues}

\begin{table}[h]
 \center
 \caption{Numbers of clusters at various redshifts}
 \label{tab:totnumbers}
\begin{tabular}{llrrrr}
\hline
      &          & \multicolumn{4}{c}{Redshift}\\
Model & Relation & 0.0 & 0.5 & 1.0 & 1.5\\
\hline
{\it Radiative} & Total                     & 340 & 190 & 85 & 31\\
                & $T_{\rm dyn}-M_{500}$     & 330 & 186 & 84 & 31\\
                & $T_{\rm gas}-M_{500}$     & 332 & 186 & 82 & 31\\
                & $T_{\rm bol}-M_{500}$     & 319 & 151 & 64 & 24\\
                & $L_{\rm bol}-M_{500}$     & 317 & 186 & 85 & 31\\
                & $L_{\rm bol}-T_{\rm bol}$ & 256 & 95  & 34 & 14\\
\hline
{\it Preheating} & Total                     & 283 & 147 & 59 & 22\\
                & $T_{\rm dyn}-M_{500}$      & 273 & 143 & 56 & 22\\
                & $T_{\rm gas}-M_{500}$      & 271 & 143 & 56 & 22\\
                & $T_{\rm bol}-M_{500}$      & 264 & 134 & 53 & 22\\
                & $L_{\rm bol}-M_{500}$      & 269 & 143 & 59 & 22\\
                & $L_{\rm bol}-T_{\rm bol}$  & 190 & 92  & 48 & 14\\
\hline
{\it Feedback} & Total                      & 342 & 98  & 45 & 13\\
                & $T_{\rm dyn}-M_{500}$     & 328 & 96  & 43 & 12\\
                & $T_{\rm gas}-M_{500}$     & 327 & 89  & 41 & 11\\
                & $T_{\rm bol}-M_{500}$     & 305 & 90  & 39 & 10\\
                & $L_{\rm bol}-M_{500}$     & 339 & 98  & 45 & 13\\
                & $L_{\rm bol}-T_{\rm bol}$ & 269 & 67  & 32 & 12\\
\hline
\end{tabular}
\label{tab:clusters}
\end{table}

Table~\ref{tab:clusters} lists the numbers of clusters in our
catalogues for each of the simulations at all 4 redshifts. The first
row for each model gives the total number of clusters in our
catalogues, down to a minimum mass, $M_{\rm 500}=1.2\times
10^{13}\hMsun$, corresponding to $\sim 500$ dark matter particles in
the {\it Radiative} and {\it Preheating} simulations, and $\sim 1400$
dark matter particles in the (higher resolution) {\it Feedback}
simulation. At $z=0$, each model contains around 300 clusters above
our mass limit, decreasing by around an order of magnitude by $z=1.5$.

We also made a number of additional cuts to the catalogues, specific
to each scaling relation. Firstly, we noted a small number of systems
that were significantly offset from the mean relation. On inspection,
such objects were found to be erroneous as they were subclumps falling 
into neighbouring clusters. Thus, for each relation, we
discarded all objects with $\Delta \log (Y)>0.1$, larger than intrinsic
scatter in the $T_{\rm dyn}-M_{\rm 500}$, $T_{\rm gas}-M_{\rm 500}$ and 
${T_{\rm bol}-M_{\rm 500}}$ relations; and $\Delta \log (Y)>0.5$ in the
$L_{\rm bol}-M_{500}$ and $L_{\rm bol}-T_{\rm bol}$ relations, respectively. 
Secondly, for the $L_{\rm bol}-T_{\rm
bol}$ relation, we made an additional cut in temperature, such that
the catalogues were complete in $T_{\rm bol}$ (excluding those
clusters classed as outliers). For the
{\it Radiative} model, the minimum temperatures are $kT_{\rm
bol,min}=[0.74,1.0,1.25,1.35]$ keV; for the {\it Preheating} model,
$kT_{\rm bol,min}=[0.70,0.96,1.1,1.37]$ keV; and for the {\it
Feedback} model, $kT_{\rm bol,min}=[0.59,1.12,1.31,1.58]$ keV, for
$z=[0,0.5,1,1.5]$.  The numbers of clusters remaining in each of the
relations after these cuts are also listed in Table~\ref{tab:clusters}.

\section{Results}
\label{sec:results}

\subsection{Scaling relations at redshift zero}

\begin{table}
\center
\caption{Best-fit scaling relations}
\label{tab:evoparams}
\begin{tabular}{llcccc}
\hline
Relation & Model & $\alpha$ & $\log C_0(0)$ & $\log Y_0$ & $A$ \\
\hline
$T_{\rm dyn}-M_{\rm 500}$  & {\it Radiative}  & 0.70 & 0.34 & 0.34 & 1.1 \\
                           & {\it Preheating} & 0.70 & 0.33 & 0.33 & 1.1 \\
                           & {\it Feedback}   & 0.69 & 0.33 & 0.33 & 1.2 \\
                                                                                                                           
$T_{\rm gas}-M_{\rm 500}$  & {\it Radiative}  & 0.61 & 0.33 & 0.33 & 0.9 \\
                           & {\it Preheating} & 0.61 & 0.35 & 0.35 & 0.9 \\
                           & {\it Feedback}   & 0.61 & 0.35 & 0.35 & 1.1 \\
                                                                                                                           
$T_{\rm bol}-M_{\rm 500}$  & {\it Radiative}  & 0.59 & 0.38 & 0.37 & 0.5 \\
                           & {\it Preheating} & 0.61 & 0.35 & 0.35 & 0.8 \\
                           & {\it Feedback}   & 0.64 & 0.33 & 0.33 & 1.2 \\
                                                                                                                           
$L_{\rm bol}-M_{\rm 500}$  & {\it Radiative}  & 1.82 & 1.36 & 1.36 & 3.9 \\
                           & {\it Preheating} & 1.92 & 1.40 & 1.39 & 3.1 \\
                           & {\it Feedback}   & 2.10 & 1.40 & 1.40 & 3.2 \\
                                                                                                                   
$L_{\rm bol}-T_{\rm bol}$  & {\it Radiative}  & 3.06 & 0.19 & 0.20 & 1.9 \\
                           & {\it Preheating} & 3.05 & 0.26 & 0.24 & 0.7 \\
                           & {\it Feedback}   & 3.13 & 0.28 & 0.28 & -0.6\\
\hline
\end{tabular}
\label{tab:bestfit}
\end{table}

We first present the scaling relations at $z=0$ as they will form the
basis for measuring evolution in the cluster properties with redshift.
The parameters $\alpha$ and $C_0(0)$ listed in Table~\ref{tab:bestfit}
are determined from the best least-squares fit to the relation
\begin{equation}
\log Y=\log C_0(0) + \alpha\log X,
\end{equation}
where $X$ and $Y$ represent the appropriate data sets in units of $10^{14}
\hMsun$, 1\,keV and $10^{42}\,h^{-2}\,{\rm erg}\,{\rm s^{-1}}$ for
mass, temperature and luminosity, respectively.  We will consider each
relation in turn.

\begin{figure}
\centerline{\epsfig{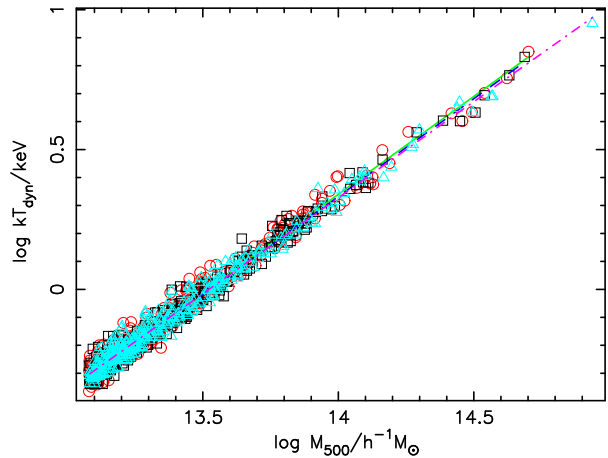}}
\caption{The dynamical temperature-mass relation within $R_{\rm 500}$ at 
$z=0$. {\it Radiative} clusters are plotted as circles (with the solid
line denoting the best-fit relation), {\it Preheating} as squares
(dashed line) and {\it Feedback} as triangles (dash-dotted line)
respectively.}
\label{fig:tdynmz0}
\end{figure}

Figure~\ref{fig:tdynmz0} illustrates the $T_{\rm dyn}-M_{500}$
relation for each of the three simulations at $z=0$, with best-fit
relations overplotted as straight lines.  The dynamical temperature is
dominated by the contribution from the more massive dark matter
particles, and so the resulting three relations are almost
identical. The measured slope of the relation is $\alpha \sim 0.7$
(Table~\ref{tab:bestfit}), close to, but slightly larger than the
self-similar value ($\alpha=2/3$); this deviation is due to the
variation of concentration with cluster mass.  When the mass-weighted
temperature of hot gas is used instead, the relation becomes flatter
than the self-similar prediction, with $\alpha \sim 0.6$. This is
expected as the excess entropy generation due to cooling and heating
is more effective in lower mass clusters (MTKP02).

\begin{figure}
\centerline{\epsfig{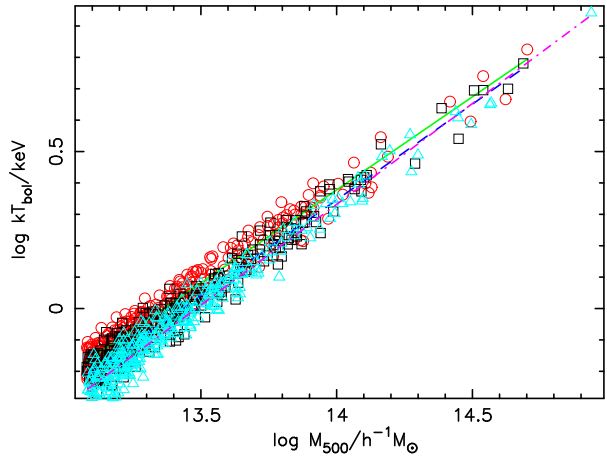}}
\caption{The bolometric emission-weighted temperature-mass relation 
within $R_{\rm 500}$ at $z=0$ for clusters in the 3 simulations. 
Symbols and lines are as in Figure~\ref{fig:tdynmz0}.} 
\label{fig:tbolmz0}
\end{figure}

Shown in Figure~\ref{fig:tbolmz0} is the $T_{\rm bol}-M_{500}$
relation for each of the 3 models.  Cool, dense gas dominates
$T_{\rm bol}$ and so this temperature is more susceptible to
fluctuations caused by merging substructure, leading to an increase in
the scatter when compared to Figure~\ref{fig:tdynmz0}.  Again, the
slope is flatter than the self-similar prediction, due to the effects
of excess entropy.  Differences between the models are larger than for
the dynamical temperature but are less than the intrinsic scatter.

\begin{figure}
\centerline{\epsfig{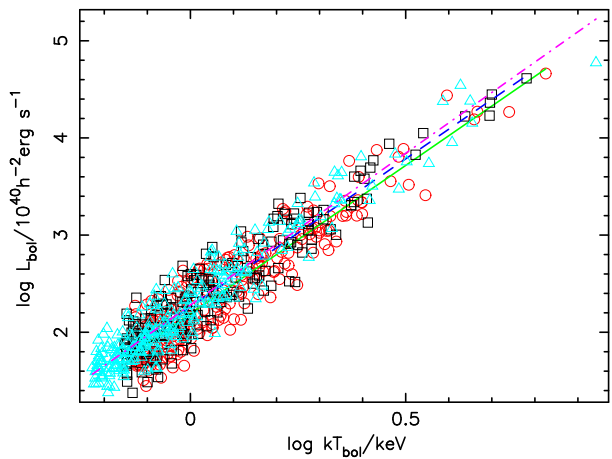}}
\caption{The bolometric luminosity-temperature relation 
within $R_{\rm 500}$ at $z=0$ of clusters
in the 3 simulations. Symbols and lines are as in Figure~\ref{fig:tdynmz0}.}
\label{fig:lbolt0}
\end{figure}

Finally, we consider relations involving the bolometric luminosity of
the cluster.  Fitting the relation between luminosity and mass, we
find a slope in the range $\alpha \sim 1.8-2.1$, significantly steeper
than the self-similar prediction ($\alpha=4/3$).  The departure from
self-similarity is exacerbated when we plot bolometric luminosity
against temperature (Figure~\ref{fig:lbolt0}).  Here, $\alpha \sim 3.1$
in all models, compared to $\alpha=2$ for the self-similar case.  The
$L_{\rm bol}-T_{\rm bol}$ relations from the three simulations are in
reasonable agreement with one another and in good agreement with the
observed luminosity--temperature relation (see MTKP02,KTJP04).

In summary, all three models successfully generate excess entropy in
order to break self-similarity at the level required by the
observations at low redshift ($z \sim 0$).  Thus, based on the local
scaling relations alone, we cannot easily discriminate between the
source of the entropy excess in clusters: whether it is mainly due
to radiative cooling, additional uniform heating at high redshift
(prior to cluster formation) or localized heating from galaxy
formation at all redshifts.

\subsection{Evolution of scaling relations with redshift}

\begin{figure}
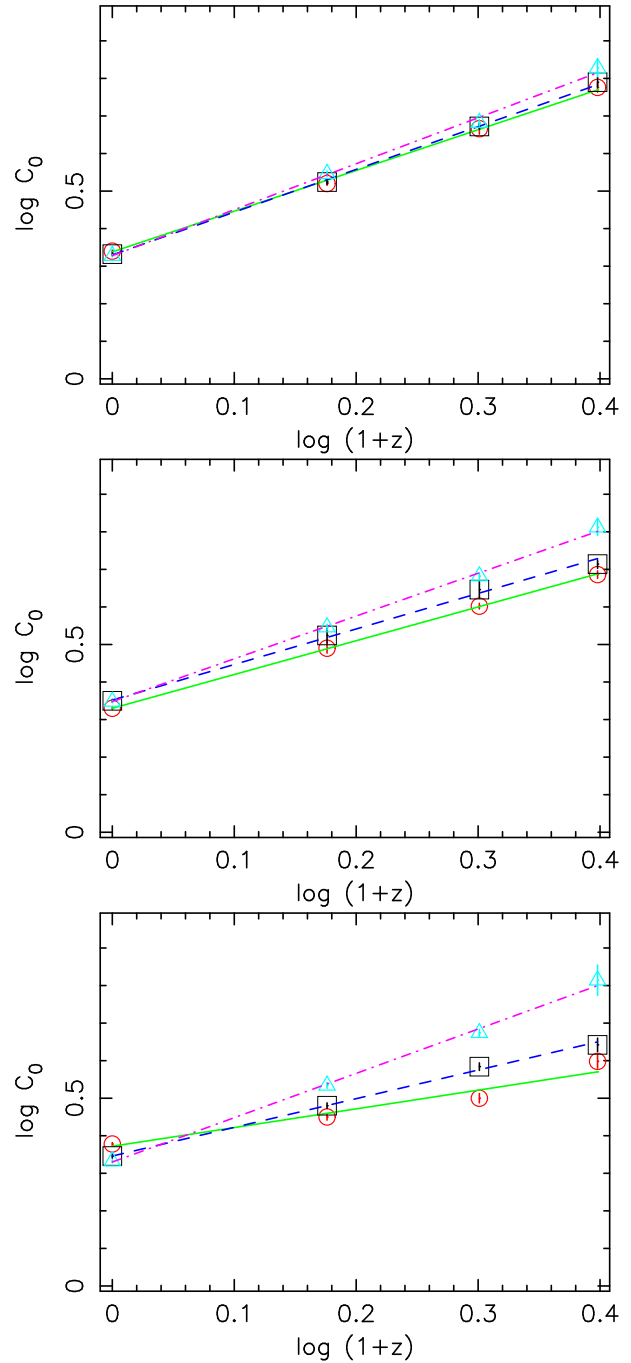

\centerline{\epsfig{file=f5a.ps,width=6cm,angle=270}}
\centerline{\epsfig{file=f5b.ps,width=6cm,angle=270}}
\centerline{\epsfig{file=f5c.ps,width=6cm,angle=270}}
\caption{ The normalization of the various temperature-mass relations as a
  function of redshift, for clusters in the {\it Radiative} (solid
  line), {\it Preheating} (dashed line) and {\it Feedback} (dot-dashed
  line) simulations: dynamical temperature (top panel), mass-weighted gas
  temperature (middle panel) and bolometric, emission-weighted temperature
  (bottom panel).}
\label{fig:tmevol}
\end{figure}

We now examine whether this degeneracy between models in the scaling
relations at $z=0$ can be broken by examining the cluster population
at higher redshifts ($z=0.5, 1, 1.5$).  None of the relations require a
significant variation in $\alpha$ with redshift.  To make our results
easier to interpret, therefore, we use simple power-law relations of
the form given in equation~\ref{eqn:powerlaw} with $\alpha$ fixed at
the $z=0$ values given in Table~\ref{tab:bestfit}.

To find the evolution of each relation, we first determine the
normalizations, $C_0$, and their corresponding error bars, 
at each redshift in the same manner as
described for redshift zero in Section~4.1 above. We then
minimize the $\chi-$squared to obtain parameters
$Y_0$ and $A$ as listed in Table~\ref{tab:bestfit} to fit
the relation
\begin{equation}
\log C_0 = \log Y_0 + A\log(1+z).
\label{eqn:normz}
\end{equation}

\subsubsection{Temperature-Mass Evolution}

In Figure~\ref{fig:tmevol}, we present values of $\log(C_0)$ versus
redshift for the three temperature--mass relations, with the best-fit
straight line overplotted.  For the $T_{\rm dyn}-M_{500}$ relation
(upper panel), we find similar evolution parameters for the three
models, $A=1.1-1.2$, confirming that including the effects of baryonic
physics does not significantly affect cluster dynamics.  The slight
excess over the self-similar value of $A=1$ is consistent with the
changing cluster concentrations.

However, both the mass-weighted temperature (middle panel) and
especially the emission-weighted temperature (lower panel) show
significant variation between the three models.  In each case the {\it
Feedback} simulation approximately follows the scaling found for the
dynamical temperature, with the {\it Preheating} and the {\it
Radiative} simulations showing progressively larger deviations below
the expected normalization as the redshift increases.  The explanation
for this lies in the variation of temperature of gas particles within
each cluster and how this changes with redshift in the different
models.  We shall explore this further in Section~\ref{sec:discuss}.

\subsubsection{Luminosity-Mass Evolution}

\begin{figure}
\centerline{\epsfig{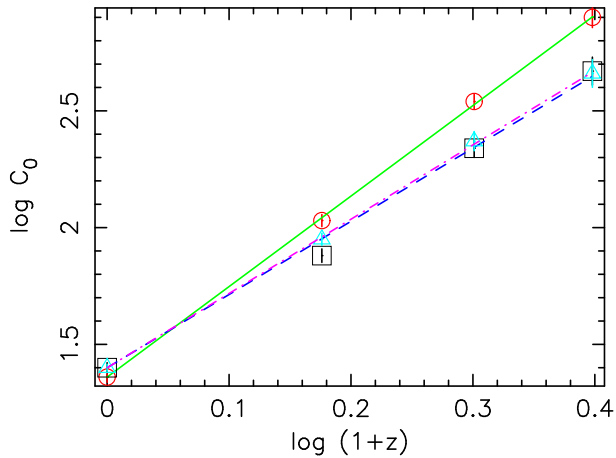}}
\caption{The normalization of the bolometric luminosity--mass
relation as a function of redshift, for clusters in the
{\it Radiative} (solid line), {\it Preheating} (dashed) and
{\it Feedback} (dot-dashed) simulations.}
\label{fig:lmevol}
\end{figure}

Figure~\ref{fig:lmevol} illustrates the normalization of the $L_{\rm
bol}-M_{500}$ relation versus redshift for all three models.  The {\it
Preheating} and {\it Feedback} models evolve almost identically with
redshift ($A \sim 3$), but the {\it Radiative} run evolves more
strongly ($A \sim 4$).  These bracket the self-similar value, $A=3.5$,
however, this agreement is somewhat coincidental because the slope of
the relation at fixed redshift is much steeper than expected
($\alpha\sim1.8$--2.1 rather than 1.3). The reason why the
{\it Radiative} simulation has steeper evolution is because of
enhanced emission from cool gas at high redshift relative that that at
low redshift---see discussion in Section~\ref{sec:discuss}.

\begin{figure}
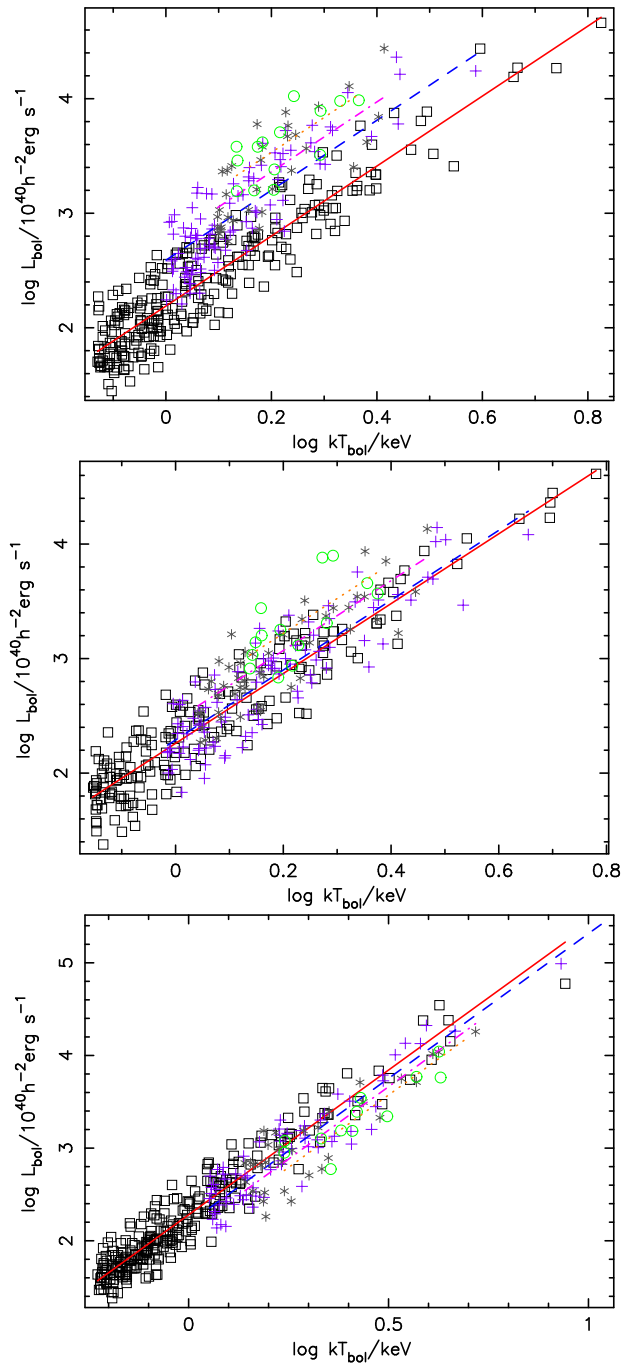

\centerline{\epsfig{file=f7a.ps,width=6cm,angle=270}}
\centerline{\epsfig{file=f7b.ps,width=6cm,angle=270}}
\centerline{\epsfig{file=f7c.ps,width=6cm,angle=270}}
\caption{The bolometric luminosity--temperature relation 
for clusters at $z=0,0.5,1,1.5$ 
(solid,dashed,dot-dashed,dotted lines respectively).
various redshifts. The top panel
is for the {\it Radiative} simulation, the middle for the
{\it Preheating} simulation and the bottom for the {\it Feedback}
simulation.}
\label{fig:lbolt4z}
\end{figure}

\subsubsection{Luminosity-Temperature Evolution}

\begin{figure}
\centerline{\epsfig{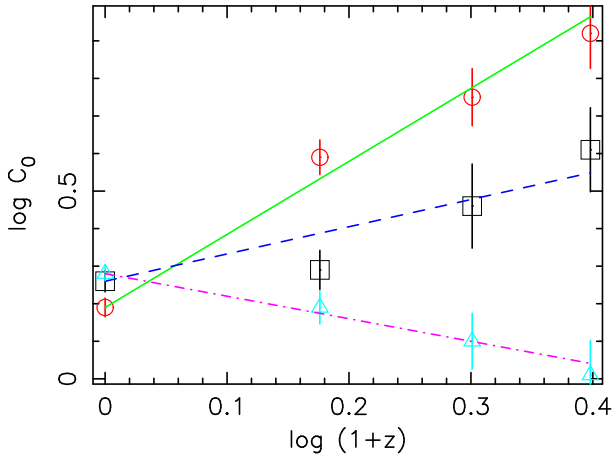}}
\caption{The normalization of the bolometric luminosity--temperature
relation as a function of redshift, for clusters in the
{\it Radiative} (solid line), {\it Preheating} (dashed) and
{\it Feedback} (dot-dashed) simulations.}
\label{fig:ltevol}
\end{figure}

Finally, we consider the evolution of the $L_{\rm bol}-T_{\rm bol}$
relation, with the relations at each redshift shown explicitly for
each model in Figure~\ref{fig:lbolt4z}, and the variation of
normalization with redshift illustrated in Figure~\ref{fig:ltevol}.
It is interesting to note that the values of $A$ are significantly
different between all three models: the {\it Feedback} model predicts
mildly negative evolution ($A=-0.6$), the {\it Preheating} mildly
positive evolution ($A=0.7$) and the {\it Radiative} strongly positive
evolution ($A=1.9$).  The latter two models straddle the self-similar
value ($A=1.5$).

The difference in slopes between the {\it Feedback} and {\it Preheating}\
runs is driven by the differences in their temperature.  The further
difference between the {\it Preheating} and {\it Radiative} runs comes
roughly equally from the temperature and luminosity evolution.

\section{Discussion}
\label{sec:discuss}

In this paper, we have focused on the evolution of cluster scaling
relations in three simulations, each adopting a different model for
non-gravitational processes that affect the intracluster gas. In the
first, {\it Radiative} model, the gas could cool radiatively and a
significant fraction cooled down to low temperatures and formed stars
(MTKP02). In the {\it Preheating} model, the same was true, although
the gas was additionally heated uniformly and impulsively by 1.5 keV
per particle at $z=4$, before cluster formation.  In the third, {\it
Feedback} model, the heating rate was local and quasi-continuous, in
proportion to the star-formation rate from cooled gas. All three models are
able to generate the required level of excess core entropy in order to
reproduce the $L_{\rm X}-T_{\rm X}$ relation at $z=0$ (MTKP02, KTJP04).

The most striking result presented in this paper is that the three
models predict widely different $L_{\rm X}-T_{\rm X}$ relations at
high redshifts. The {\it Radiative} model predicts strongly positive
evolution, the {\it Preheating} model mildly positive evolution and
the {\it Feedback} model, mildly negative evolution.  At this point,
it should be stressed that the values of $A$ presented in
Table~\ref{tab:bestfit} should not be taken too seriously.  No attempt
has been made to convert the bolometric, emission-weighted fluxes used
in this paper to observable X-ray fluxes in different instrumental
bands.  Also, the volume of our simulation boxes is such that we only
get a modest number of relatively poor clusters at high redshift.
Nevertheless, the qualitative difference between the models is very
encouraging and suggests that evolution of X-ray properties may act as
a strong discriminant between models in the future.

Previous work \citep[e.g.~][]{Pearce00,Muanwong01} has made great play
of the fact that radiative cooling can remove low-entropy material and
lead to a raising of the gas temperature above the virial temperature
of the host halo.  That effect is reproduced by the {\it Radiative}
simulation in this paper, but it is interesting to note that the
bolometric X-ray temperature exceeds the dynamical temperature of the
clusters only at very low redshift, $z\approxlt0.1$.  At higher
redshifts it falls below the dynamical temperature and is a factor of
1.6 lower by $z=1.5$.  This departure from self-similarity is a
consequence of the changing density parameter, $\Omega$, in the
concordance \LCDM\ cosmology: at high redshift $\Omega$ is close to
unity and structures grow freely; at lower redshifts $\Omega$ falls
well below unity and the rate of growth of cosmic structures declines.

The behavior of the {\it Preheating} simulation is similar, although
the relative decline in the ratio of the gas to the dynamical
temperature is smaller.  The effect cannot, therefore, be due to the
cooling of intracluster gas in the cluster cores between a redshift of
1.5 and the present---in the {\it Preheating} run very little gas
cools below a redshift of 4 and so that cool core gas would still be
there today.  Instead, we attribute the presence of cool gas to the
accretion of low temperature subclumps.  Such accretion is a
ubiquitous feature of clusters \citep[e.g.~][]{Rowley04}.  Cool gas is
seen in maps of clusters at low redshift \citep{Onuora03} and would be
expected to be much more prevalent in clusters at high redshift in the
\LCDM\ cosmology.

To test this hypothesis, we measured the temperature variation within
each cluster as follows.  First we averaged properties within 20
spherical annuli out to $R_{500}$, to create smoothed dynamical,
$\bar{T}_{\rm dyn}$, and gas temperature, $\bar{T}_{\rm gas}$,
profiles.  Then we measured the mean deviation (in log space) of the
gas temperature from the local dynamical temperature
\begin{equation}
\tau={1\over N}\,\sum_i\left(\log_{10}T_i-\log_{10}\bar{T}_{\rm dyn}\right),
\end{equation}
and the root-mean square deviation of the temperature, $\sigma_T$,
from the mean,
\begin{equation}
\sigma_T^2={1\over N}\,\sum_i\left(\log_{10}T_i-\log_{10}\bar{T}_{\rm
  gas}\right)^2,
\end{equation}
where $N=\sum_i$ and the sum runs over all hot gas particles
($T_i>10^5$\,K) within $R_{500}$.

\begin{figure}
\centerline{\epsfig{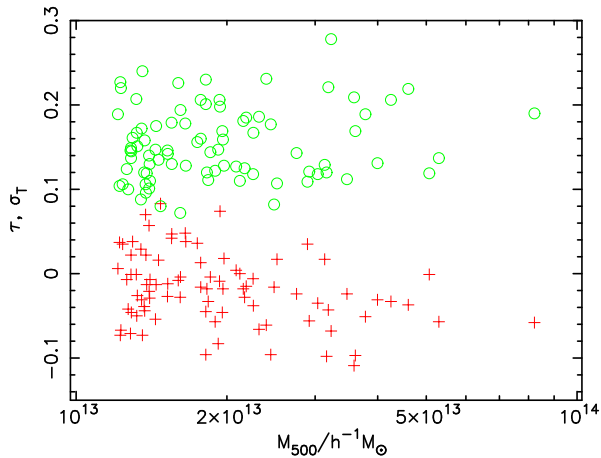}}
\caption{Values of $\tau$ (crosses) and $\sigma_T$ (circles) for
  individual clusters in the {\it Radiative} simulation at $z=1$.}
\label{fig:delt6720.1002}
\end{figure}

As an example, Figure~\ref{fig:delt6720.1002} shows the values of
$\tau$ (crosses) and $\sigma_T$ (circles) for each cluster at $z=1$ in
the {\it Radiative} simulation.  This particular example has been
chosen simply because the two properties are well-separated and easy
to distinguish on the plot.  As can be seen, the mean gas temperature
of the more massive clusters typically lies below the local dynamical
temperature, by as much as 15 per cent at this redshift.  However, the
dispersion in temperature is much larger, typically a factor of 1.5,
so there will be fluctuations both above and below the dynamical
temperature.

\begin{figure}
\centerline{\epsfig{file=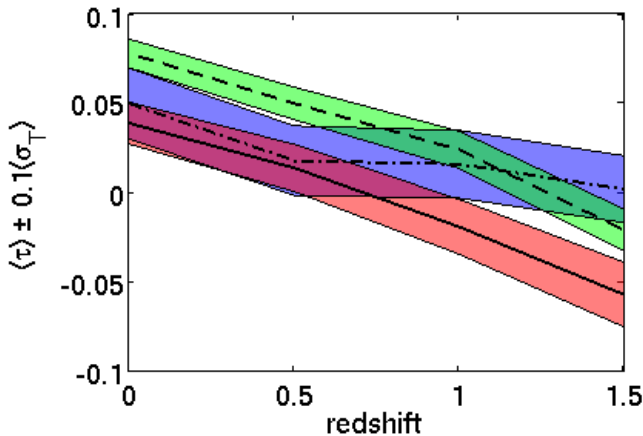,width=8.7cm}}
\caption{The mean value of $\tau$ for clusters in the {\it Radiative}
  (solid line), {\it Preheating} (dashed line) and {\it Feedback}
  (dash-dotted line) simulations as a function of redshift.  The half-width
  of the shaded regions corresponds to the mean value of $\sigma_T$ at
  that redshift.  All averages are for clusters with mass greater than
  1.2$\times10^{13}\hMsun$.}
\label{fig:delt}
\end{figure}

Figure~\ref{fig:delt} demonstrates visually the evolution of $\tau$
and $\sigma_T$ with redshift.  At each redshift, the plot shoes the
average value of $\tau$ over all clusters with masses greater than
1.2$\times10^{13}h^{-1}$\Msun.  The half-width of the shaded regions
represent the average values of $\sigma_T$, divided by 10 for clarity.
Concentrating first on the {\it Radiative} simulation, it can be seen
that the mean cluster temperature increases relative to the virial
temperature over time and that the dispersion in temperature
decreases.  This is consistent with a decreasing amount of
substructure within the clusters at lower redshift, although it should
be noted that part of the effect is due to the narrower range of
cluster masses resolved by the simulations at high redshift, as the
average value of $\tau$ decreases with increasing cluster mass.

The behavior of the {\it Preheating} simulation mimics that of the
{\it Radiative} one, but with a bias to higher mean temperatures.  The
{\it Feedback} simulation, however, is quite different.  It shows a
much larger dispersion than the other two, but no bias to low
temperatures at high redshift.  That is because gas is free to cool
down to low temperatures but some of that gas is then heated back up
high temperatures by the feedback.  At low redshift cooling becomes
less important and the {\it Feedback} run then shows a slight rise in
$\tau$, matching that seen in the other two runs.

Our results have two very important implications for observations of
high redshift clusters.  Firstly, the behavior of the $L_{\rm
X}-T_{\rm X}$ relation at high redshift will determine the number of
high redshift clusters to be found in surveys such as the ongoing {\it
XMM-Newton} Cluster Survey \citep{Romer01} and this will have a
significant impact upon their use in probing cosmological parameters.
A positive evolution such as that shown by the {\it Radiative}
simulation will yield many more observable high-redshift clusters than the
negative evolution of the {\it Feedback} model.  As discussed in
Section~\ref{sec:srel} and summarized in Figure~\ref{fig:ltevolobs},
the observational situation is far from clear but does seem to indicate
positive evolution.

Turning this argument around, our results suggest that observational
constraints on the degree of evolution of the $L_{\rm X}-T_{\rm X}$
relation will allow interesting constraints to be placed on the source
of entropy generation in clusters, in particular the relative role of
cooling and heating and whether most of the heating of the
intracluster gas occurred at high redshift (as in the {\it Preheating}
model) or was a continuous function of redshift (as in the {\it
Feedback} model).  Taking our results at face value with recent
observations would suggest that our {\it Feedback} model is generating
too much excess entropy at $z<1.5$ and that the bulk of the heating must
have occurred at higher redshift.  However, we stress once again that
this result is very tentative.

\section{Conclusions}
\label{sec:conclude}

The evolution of X-ray cluster scaling relations are a crucial
component when constraining cosmological parameters with
clusters. Observational studies at low redshift have already shown
that the scaling relations deviate from self-similar expectations,
attributed to non-gravitational heating and cooling processes, but
their redshift dependence is only starting to be explored. In this
paper we have investigated the sensitivity of the X-ray scaling
relations to the nature of heating processes, using three numerical
simulations of the $\LCDM$\ cosmology with different heating
models. While all three simulations reproduce more or less the same
scaling relations at $z=0$ (as they were designed to produce the
correct level of excess entropy), they predict significantly different
results for the evolution of the \Lx-\Tx\ relation to $z=1.5$.

In conclusion, our findings strongly suggest that the relative
abundance of high and low redshift clusters will place interesting
constraints on the nature of non-gravitational entropy generation in
clusters.  First indications are that an early and widespread
preheating of the ICM is to be preferred to an extended period of
preheating that is associated with galaxy formation.  However much
more detailed modeling is required and the observational picture is
as yet unclear.

\acknowledgements The simulations described in this paper were carried
out on the Cray-T3E at the Edinburgh Parallel Computing Centre and the
COSmology MAchine in Durham as part of the Virgo Consortium
investigations into the formation of structure in the Universe. OM is
grateful for the hospitality and support of the Astronomy Centre at
the University of Sussex where much of work in this paper was carried
out, and for financial support from Khon Kaen University. She and PAT
also acknowledge support from the Thailand Research Fund and the
Commission on Higher Education grant MRG4680129.


\end{document}